\documentclass[showpacs,preprintnumbers,amsmath,amssymb,nofootinbib]{revtex4}
\usepackage{graphicx}
\usepackage{epsfig}
\usepackage{bm}
\usepackage{amsfonts}
\newcommand{\Po}{\cal P}
\newcommand{\erfc}{\text{erfc}}
\newcommand{\eq}{\begin{eqnarray}}
\newcommand{\eqx}{\end{eqnarray}}
\newcommand{\ba}{\begin{equation}}
\newcommand{\ea}{\end{equation}}
\newcommand{\f}[2]{\frac{#1}{#2}}

\newcommand{\nn}{{\cal N}}

\newcommand{\dl}{\delta}
\newcommand{\sig}{\sigma}
\newcommand{\pa}{\partial}

\newcommand{\cor}[1]{\left\langle{#1}\right\rangle}

\newcommand{\bit}{\begin{itemize}} 
\newcommand{\eit}{\end{itemize}}

\def\tr{\tilde r}
\def\la{\label}
\def\nn{\nonumber \\}
\def\bi{\bibitem}
\def\g{\gamma}

\def\ka{\kappa}
\def\lam{\lambda}
\def\al{\alpha}
\def\alp{\alpha'}

\def\va{\varphi}
\def\bv{\bar \varphi}
\def\eps{\epsilon}
\begin{document}

\title{Traveling wave solution of the Reggeon Field Theory}

\author{Robi Peschanski }
\email{robi.peschanski@cea.fr} \affiliation{Institut de Physique 
Th\'eorique,\\
CEA, IPhT, F-91191 Gif-sur-Yvette, France\\
CNRS, URA 2306 }

\begin{abstract}
We identify  the nonlinear evolution equation  in impact-parameter space for 
the ``Supercritical Pomeron'' in Reggeon Field Theory  as  a 2-dimensional 
stochastic Fisher-Kolmogorov-Petrovski-Piscounov equation. It exactly  preserves unitarity and leads in its radial form  to an high energy traveling wave  solution corresponding to an ``universal'' behaviour of the impact-parameter front profile of the elastic amplitude; Its rapidity dependence and  form depend only on one parameter, the noise strength, independently of the initial conditions and of the non-linear terms restoring unitarity.  Theoretical predictions are presented for the three typical distinct regimes corresponding to zero, weak and strong noise.
\end{abstract}

%\pacs{} 
\maketitle

\section{Introduction}
\la{Introd}
The question of the high-energy behavior of soft hadron-hadron amplitudes 
and 
in particular of their expanding impact-parameter disk with rapidity is a rather 
old subject, but still not solved, being in the basically unknown realm of non-perturbative QCD. 
However, one  promising theoretical approach  at early time is the 
Reggeon Calculus \cite{gribov} and, in the formalism which we will be dealing with in the present work the Reggeon
Field Theory (RFT) \cite{abarbanel},
where the amplitude is described in terms of  an effective quantum field 
theory of ``Pomeron fields''. It  derives from a Lagrangean
in $2\!+\!1$ dimensions, where ``space'' is the 2-d impact-parameter $\vec 
b$ 
and ``time'', the overall rapidity $Y.$ In the studie performed during the 70's, and after a series of works dedicated to the renormalization group approach to the RFT  \cite{abarbanel}, it appeared also 
\cite{supercritical,ciafaloni,parisi} that a 
physically interesting case is when one considers instead a 
``supercritical'' 
bare Pomeron
$P$, $i.e.$  when the intercept $\al$ is greater than 1 (in fact
$\al>\al_c>1,$ where $\al_c$ includes the quantum effects of the renormalizable RFT when the Pomeron field is  
at criticality \cite{abarbanel}). In that case, unitarity is violated by 
the 
 bare Pomeron (equivalent to a Born term) but expected to be recovered thanks to damping Pomeron interactions. 
It was  shown that the 
impact-parameter 
disk was expanding like the rapidity $Y,$ expressing a dynamical instability 
ofthe RFT \cite{parisi}. However, the field theoretical techniques known at that 
time did not seem to give much more indication on the solutions. 

The notion of a Reggeon Field Theory and its use to describe ``soft'' hadronic interactions appeared, since that time  and until recently, under various forms which may differ from the original version depicted in \cite{abarbanel}. For instance, they may refer to the similar approach  deduced or inspired by QCD and its dipole model \cite{Iancu,QCD}. Also, there exists a  phenomenological interest for using a ``supercritical Pomeron'' in models based on interacting Pomerons \cite{models}. So, we want to specify now in which sense we use the RFT and what is different in our approach from the previous ones. First, we want to address the problem of finding the solutions of the full 2-dimensional transverse space problem including an explicit form of the impact-parameter dependence of the elastic amplitude. To our knowledge, explicit solutions have been only found  in the zero-dimensional approximation only. Two-dimensional formulations  of the QCD Pomeron calculus in the dipole approach are also widely discussed \cite{Iancu,QCD}, but  explicit solutions seem to be difficult to acquire. So, we restrict our analysis to the initial formulation \cite{abarbanel}  and thus our starting point for the Lagrangean is the original one \cite{abarbanel}. Hence, when we use the term Reggeon Field Theory (RFT) we refer in the present paper to that precise formulation, up to a suitable generalization to be discussed further on.

The goal of our paper is to update 
the original  study of RFT by introducing  new powerful tools known under the name of ``traveling wave solutions'' of non-linear evolution equations and already used in a different context for QCD evolution equations \cite{munier}. In fact one of our motivations is to try and give a theoretical  answer to an old question raised by the phenomenology of elastic hadronic reactions, and its approach by interacting supercritical Pomerons. The  ``soft'' hadronic elastic amplitudes seem to follow a common  behaviour at high-energy, independent of the reaction one considers. While this property could be understood by the factorization properties of a single Regge pole exchange (see, $e.g.$ \cite{donach}), it is not known how a property may be obtained from an interacting Pomeron framework where the ``bare'' Pomeron input is  deeply modified by the interactions.

The relation of RFT with non-linear evolution equations (which will allow to use the traveling wave framework) has appeared  since long in relation with statistical mechanics of  out-of equilibrium processes. Starting with the deep relation between the original RFT with directed percolation \cite{critical}, there is a long list of works using this kind of connection , in particular with stochastic evolution equation of Langevin type. In fact some of the works (see, $e.g.$ \cite{QCD,Iancu}) are using this connection to try and derive the stochastic evolution equation corresponding to QCD Reggeon Calculus. We shall indeed use some tools from statistical mechanics, in particular those developed in Ref. \cite{langevin} to transform the RFT formulation in terms of a Langevin equation of known type and possessing traveling wave solutions.

The main feature of traveling wave solutions of non-linear evolution equations is that they lead  to ``universality'' 
properties, that are properties which will be of general value, $i.e.$ 
irrespective of particular initial conditions or on features of the equation such as  the form of the non-linear damping terms responsible for the unitarity restoration. Our hope is thus to provide through the traveling wave approach, an explicit high-energy solution of the RFT and in the same footing 
a physical understanding of the empirical ``universal'' properties of soft scattering amplitudes at high energies which are difficult to 
explain in a supercritical Pomeron framework. Our paper thus contains the  theoretical derivation of the solutions of the (2+1) dimensional RFT \cite{abarbanel} $via$ the identification of the related stochastic non-linear Langevin equations.

The paper is organized as follows. In section \ref{RFT}, we show that the RFT, with a supercritical bare Pomeron as  input, can be 
found equivalently realized by the 2-dimensional version of the stochastic Fisher and Kolmogorov, Piscounov, Petrovsky (sFKPP) equation, 
and that the elastic amplitude is  solution of its reduction to the 1-dimensional radial (azimuthally symmetric) form. In 
section \ref{model}, we derive the main feature of the  mean field (or deterministic) radial FKPP equation: the existence and universal 
properties of circular traveling wave asymptotic solutions. In section \ref{stoc} we introduce the effect of stochasticity by 
analyzing the solution dependence on the noise term. It  appears with two markedly different regimes at weak and strong noise strengths. In section \ref{out} we present our conclusions and an outlook on the theoretical  implications of the traveling wave picture. In the appendices we show the derivation of the solution in the deterministic radial case and an overview on possible phenomenological implications.
\section{From Reggeon field theory to the 2-d sFKPP equation}
\la{RFT}

The RFT with a supercritical Pomeron is defined \cite{abarbanel} from the following ingredients, namely one propagator $\Po 
\to 
\Po,$ with coupling $\mu$ corresponding to the bare supercritical Pomeron intercept 
$1\!+\!\mu> 
1.$ There is a kinetic term in impact-parameter space 
with coupling identified with $\alp,$ the slope of the bare Pomeron 
trajectory. For the 
Pomeron interaction vertices, one includes  the
triple Pomeron vertex, which gives rise to two possible contributions, $i.e.$ the $merging$ triple 
Reggeon term   
$\Po\!+\!\Po \to \Po$ and the  $splitting$ term $\Po \to \Po\!+\!\Po,$ 
with initially
equal strength $\lambda$ corresponding to the triple-Pomeron coupling.

The field theory action
%corresponding to Eq.\eqref{langevin}
 is defined in terms of quantum bosonic fields 
$\varphi$ and the conjugate
%their auxiliary counterparts 
$\bar \varphi$ by the action \cite{abarbanel}
\ba
S[\varphi,\bar \varphi]=\f 1\alp\ \int \!d^2b\ 
dY\left\{\bar \varphi\left[\pa_Y\!-\!\alp\nabla^2\right]\varphi\!-\!\mu\ \bar 
\varphi \varphi -i\lam\left(\bar 
\varphi
\varphi^2\!+\bar 
\varphi^2\varphi\right)\right\}\ .
%\!-\!
%\kappa\mu\ \bar \varphi^2 (\varphi-\varphi^2)
%\right\}
\la{action0}
\ea
As discussed in
 \cite{abarbanel}, the imaginary coupling constant makes this theory non-hermitian and thus the fields $\va$ and $\bv$ do not play a symmetric role through time reversal. It is indeed convenient to perform the field transformation $\va\!\to 
i\va, \bv \to -i\bv,$ giving rise to  the modified action
\ba
S[\varphi,\bar \varphi]=\f 1\alp\ \int \!d^2b\ 
dY\left\{\bar \varphi\left[\pa_Y\!-\!\alp\nabla^2\right]\varphi\!-\!\mu\ \bar 
\varphi \varphi +\lam \bar 
\varphi
\varphi^2\!-\lam \bar 
\varphi^2\varphi\right\}\ .
%\!-\!
%\kappa\mu\ \bar \varphi^2 (\varphi-\varphi^2)
%\right\}
\la{action}
\ea
One important outcome of RFT is its remarkable connection with problems of non-equilibrium statistical physics \cite{critical}. 
Following a known technique \cite{langevin}, the fields $\bv$ can be integrated out, and play the role 
of  auxiliary fields appearing as external source fields for the deterministic part
(for linear terms in $\bar \varphi $) and the noise terms (for quadratic terms 
in $\bar \varphi$) of a non-linear Langevin equation, as we shall now see. Following \cite{langevin}, 
one linearizes the remaining quadratic $\bar 
\varphi^2$ contribution in \eqref{action} by introducing  a stochastic white noise via a Stratonovitch transformation\footnote{The linearization of the quadratic terms in the action is a well-known procedure. The interested reader will find in ref.\cite{langevin} a detailed derivation of the transformation of the action \eqref{action} leading to the Langevin formulation \eqref{langevin0}.}, in such a way that 
all terms become linear in $\bv.$ Then performing the path 
integral over   $\bar \varphi $ boils down  to a 
 a 
nonlinear Langevin equation for the field $\va$ which  now acquires the interpretation of  random realizations of the (properly normalized) elastic 
scattering amplitude $T$, namely
\ba
 \frac d{dY} \ T\left(Y,\vec b\right)= \alp\ \nabla^2_b T +\mu\ T - \lam
\ T^2  
+\sqrt{2\alp\lam\ T}\ \nu(Y,\vec b\ )\ ,
 \label{langevin0}
\ea
 where the white noise verifies 
 \ba
 \cor{\nu(Y,\vec b\ ),\nu(Y',\vec b'\ )} =\dl(Y'-Y)\dl^2(\vec b'-\vec b\ )\ 
.
 \label{noise}
\ea

We have now to introduce the appropriate normalization of the scattering amplitude in impact-parameter space which is imposed by the  
unitarity limit $T\equiv 1.$ This comes as a constraint both on the deterministic and on the stochastic part of \eqref{langevin0}, since 
$T\equiv 1$ should appear as a stable {\it fixed point} of the equation\footnote{The stable fixed point is finally reached at infinite rapidity. In fact, one could technically equivalently consider an unitarity-preserving   fixed point of \eqref{langevin0} at $T\equiv \mu/\lam \le 1.$ However, it is more often considered that the black disk limit $T\equiv 1$ is the physical one.} (the other fixed point  is the ``unstable'' fixed point at $T=0,$ since the rapidity evolution increases, at least in average, the value of $T$). The unitarity constraint thus leads us to modify
Eq.\eqref{langevin0} to get the following form
\ba
 \frac d{dY} \ T\left(Y,\vec b\right)= \alp\ \nabla^2_b T +\mu\left(T - 
T^2\right) 
+\sqrt{2\alp{\bf \ka}\mu\ (T-T^2)}\ \nu(Y,\vec b\ )\ ,
\la{langevin}
\ea
where it imposes equal coupling $\lam \equiv \mu$ to the terms in $T$ and $T^2$
of the deterministic equation and adding a $T^2$ term in the noise factor in order to ensure it to vanish at the black disk limit.

A key feature of Eq.\eqref{langevin} compared to the initial formulation \eqref{langevin0} is the introduction of the parameter $\ka$ which plays a crucial role both physically and mathematically on the nature of the solutions.
At first we note that  unitarity  imposes no {\it a priori} constraint on the noise strength and thus allows for the introduction of the parameter $\ka.$ Physically, $\ka$ introduces a parametric factor between the strength of the  $merging$ term   
$\Po\!+\!\Po \to \Po$ and the  $splitting$ term $\Po \to \Po\!+\!\Po.$ This degree of freedom may come from at least two physial motivations. First, we will see that the traveling wave solutions possess ``universal'' features, in particular they will remain valid for more complicated $merging$ factors ($e.g.\ T^n, n> 2$ or even with a positive monotonous function $T^2f(T)$ with $f(1)=1$). Hence there is  $a\ priori$  no constraint of equal coupling between $merging$  and  $splitting$ terms. A more intringuing motivation may come from an analogy with the dipole picture. Taking into account their size, the $merging$  and  $splitting$ rules for dipoles imply a size-dependence of their effective coupling strength. Indeed, ``fat'' dipoles may merge more easily than ``thin'' ones, since it requires a matching of their transverse coordinates, while splitting does not seem to require such an effect. All in all, we find it 
physically suitable to consider the generalized Eq.\eqref{langevin} as the basic equation to be solved. Obviously taking $\ka=1,$ we recover the original RFT, up to a quadratic $T^2$ term in the noise which is easy to reinterpret as a four-vertex, see further.

Our basic starting point is to point out that Eq.\eqref{langevin} is (by introducing 
canonical variables, see further)  the extension in two spatial dimensions of the Fisher and Kolmogorov, Piscounov, Petrovsky (FKPP) equations (for $\ka=0$), or (for $\ka\ne 0$) its  stochastic extension (sFKPP), see Refs.\cite{FKPP,wave,brunet,wave1,mueller,munier}. This will allow us to find the solutions of the RFT for a supercritical bare Pomeron, thanks to  modern tools\footnote{It is to be mentionned that 
a first connection between  Reggeon Field Theory and circular traveling waves applied to cluster growth
 appeared already in \cite{clusters}} applied to the old and yet unsolved RFT problem.

%Eq.\eqref{langevin} calls for comments. From the point of view of RFT,  the parameter $\ka$ allows for a  modification of  
%the relative strength of the Pomeron merging and splitting vertices.  $\ka\!=\! 1\ (resp.\ \ka\!=\!0)$ corresponding to the initial
%RFT case with equal merging and splitting strength ($resp.$ no Pomeron loops). In fact, the physical context of hadronic reactions may (and probably do) indeed lead  to more general mechanisms of restoring  unitarity, not necessarily only through  the triple Pomeron coupling. One key property of sFKPP equations 
%\cite{FKPP,wave,brunet,wave1,mueller,munier} is $universality,$ meaning that an sFKPP equation has asymptotic so will provide the solution of the original RFT.

Mathematically speaking, we are looking for solutions which are not dependent of the forms of the non-linear terms provide they ensure a stable fixed point, that is  $T=1$ in our case. $Universality$ also means that the solution is independent of the initial conditions after some ``time'' (here, rapidity) evolution interval. This defines a ``universality class'' of solutions, which will ultimately depend only of the value of $\ka,$ that is on the noise strength. If we were in the situation of statistical physics at equilibrium , we could consider $\ka$ as the order parameter of the problem. In our case it will allow to separate different regimes (but not necessarily separated by  critical points.) 
We also note that the quadratic term in the noise can be easily reinterpreted in the field theoretical framework as a $\Po\!+\!\Po \to 
\Po\!+\!\Po$ coupling in the RFT framework. This is equivalent to the  following RFT action
\ba
S[\varphi,\bar \varphi]=\f 1\alp\ \int \!d^2b\ 
dY\left\{\bar \varphi\left[\pa_Y\!-\!\alp\nabla^2\right]\varphi\!-\!\mu\ \bar 
\varphi (
\varphi\!-\varphi^2)\!-\!
\kappa\mu\ \bar \varphi^2 (\varphi-\varphi^2)
\right\}\ .
\la{actionbis}
\ea
To complete the theoretical preliminaries it is worth mentioning that from the point of view of statistical physics, it is known  
that more general Langevin 
equations  can in turn be analyzed 
in terms of a bosonic quantum field theory\footnote{Doi and Peliti, Refs. 
\cite{doi}, addressed  the related problem of mapping master equations for reaction-diffusion 
processes to a field theory action.} \cite{cyrano}. This formalism is 
particularly convenient to treat  fluctuations superimposed to mean-field 
equations. Hence both techniques coming from statistical and particle physics can be joined together to 
get a deeper understanding of the original RFT problem and find the structure of its solutions, $i.e.$ identifying its 
``universality class''.

Let us introduce now canonical variables allowing to put  \eqref{langevin} in the generic form of the sFKPP equation.
By suitable redefinitions
\begin{equation}
T(Y,\vec b) \equiv \ U(t,\vec r)\quad;\quad \mu (Y\!-Y_0)=t  \quad;\quad\sqrt{\frac \mu 
\alp}\ \vec b 
=\vec r  \quad;\quad\ \epsilon=\ \sqrt{2\mu 
\ka} \ ,
 \la{redef}
\end{equation}
Eq. \eqref{langevin} can be recast in the canonical form
\begin{equation}
 \frac d{dt} \ U\left(t,\vec r\right)= \nabla^2_r U + \ U -\ U^2 + 
\epsilon\ \sqrt{U(1-U)}\ \nu(t,\vec r)
\ ,
 \la{FKPP2}
 \end{equation}
where the only remaining dimensionless parameter defines  the normalized  noise strength $\epsilon$ as a function of 
the product of  
the ``splitting over merging'' 
factor $\ka,$ and  the ``super-criticality parameter'' $\mu.$
Eq. \eqref{FKPP2} is  the canonical form  of the    nonlinear 
stochastic Fisher-Kolmogorov-Petrovski-Piscounov (FKPP) equation. It is worthwhile to note that the time relation in \eqref{redef} is defined up to a rapidity translation  $Y\!-Y_0.$ 

To remind known properties in dimension one, the 
remarkable feature of the FKPP class of equations  
\cite{FKPP,wave,wave1,munier} is to admit asymptotic traveling wave solutions, 
$i.e.$ 
solutions which depend neither   on the initial conditions nor on the 
precise 
form of the nonlinear term at large enough evolution time. In the example of the deterministic case (without noise), we display a sketch of 
the traveling wave solutions of the 1d FKPP equation on Fig.\ref{2}.
\begin{figure}[t]
\mbox{\epsfig{figure=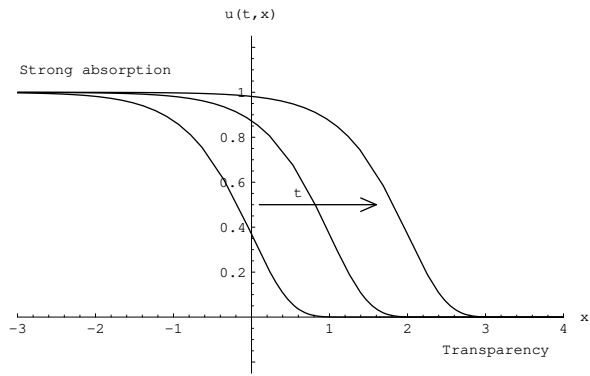,width=8.6cm,angle=0}}
\caption{{\it Traveling waves in 1 dimension}. 
The traveling waves are asymptotic solutions $u(x,t)$ of the 1-dimensional FKPP 
equation,  translation invariant in time and joining the unstable fixed 
point ({\it transparency i.e.}  dilute medium), to the stable fixed 
point ({\it strong absorption i.e.} dense medium). The figure is  from 
Ref.\cite{munier}.}
\la{2}
\end{figure}
For sFKPP, the stochastic form of 
the 
1-d equation,  following a series of applications to QCD \cite{munier,Iancu}, 
the recently found solutions \cite{mumu} can be interpreted as 
a  
stochastic superposition of traveling waves (see also \cite{beuf}).

Our aim is to look for similar properties for the 2-d version of the FKPP 
equation and find their consequences for the high-energy elastic amplitude, solution of the RFT. 
Our 
main results is the prediction of an asymptotic ``universal'' scaling form 
of 
the soft elastic amplitude $T(Y,\vec b )$ and in particular  the   prediction of an asymptotic expression for the 
expanding impact-paramter
disk.

The study of the general 2-d sFKPP equation \eqref{FKPP2} is interesting in 
itself and some results have been obtained already in the statistical physics 
literature 
($e.g.$ studies on the instabilities of the front wave \cite{kessler}). 
In the following, due to the rotational symmetry in impact-parameter space 
of 
the elastic amplitude\footnote{Note that non azimuthally symmetric 
fluctuations 
could play a role in diffractive inelastic amplitudes, or other amplitudes which are not constrained by rotation symmetry.} we shall 
concentrate 
our 
analysis of Eq.\eqref{langevin} to radial amplitudes, $i.e.$ depending 
spatially only on the radial coordinate $r=|\vec r|.$
A comment is in order at this stage. Indeed, one could consider  non azimuthally symmetric fluctuations contributing to a symmetric 
average. However, both the stochastic and the nonlinear character of the equation seems to invalidate this possibility. In particular 
anisotropic evolution may be caused by the noise \cite{kessler} and thus lead to a forbidden azimuthal symmetry breaking of the solution . We will assume that, if physical azimuthally assymetric fluctuations of the amplitude may exist, they are 
azimuthally averaged  after a characteristic time much smaller than the typical evolution time in rapidity. 

One 
then obtain, after obvious integration over the azimuth, the following equation to be 
studied
\begin{equation}
 \frac d{dt} \ U\left(t,r\right)=  \partial_{rr} \ U +\f 1r {\partial_{r}} \ U
+\ 
U -\ U^2+ \epsilon \ \sqrt{\f {U(1-U)}{2\pi r}}\ \ \nu(t,r)\ , 
 \la{radial}\end{equation}
 where a curvature term $\f 1r \partial_{r}  U$ appears in addition to the 
original 1-dimensional
FKPP equation \eqref{langevin}. Note also the modification of the noise strength by a factor $1/{2\pi r}$ reflecting the symmetry 
constraint on the fluctuation strength. The white noise satisfies $ 
\cor{\nu(t,r),\nu(t',r')} =\dl(t'\!-\!t)\dl(r'\!-\!r)$. Looking for the 
asymptotic 
universal solutions of Eq.\eqref{radial} in terms of  circular {\it 
circular} traveling 
waves in impact-parameter space is the goal of our paper.

\section{Circular traveling waves: Deterministic case}
\label{model}
Let us first consider the equation \eqref{radial} without  the  noise term, $i.e. \ 
\eps=0,$ namely
the radial 
extension of the 2-d deterministic FKPP equation. It corresponds to neglecting the Pomeron 
loop contributions
in the high-energy elastic amplitude. A series of results have been obtained 
for 
the  1-dimensional FKPP equation. The same  methods, which we will adapt for 
the 
radial case, lead to new results. Indeed, the radial case is adding the  derivative term $\f 1r\partial_{r}  U$ to the standard FKPP 
equation  and work in the half line $r \in [0,\infty]$.   The deterministic equation to be studied is thus
\begin{equation}
 \frac d{dt} \ U\left(t,r\right)=  \partial_{rr} \ U +\f 1r\ \partial_{r} \ 
U 
+\ 
U -\ U^2\ . 
 \la{radial1}\end{equation}
We will prove the 
existence of traveling wave asymptotic solutions, which appear now as {\it circular} traveling waves (picturally reminiscent of  those created by a 
stone falling in water).

Let us recall first the main guiding principles of  the FKPP traveling wave 
analysis. 
One distinguishes \cite{wave1} different regions, starting from the forward towards 
the 
backward of the wave, namely: the ``very forward'' region, the ``leading 
edge'', 
the ``wave interior'' and the ``saturation'' regions. 
The universality properties mainly pertain to the ``leading 
edge'' region and its transition to the ``wave interior''. In order to fulfill these 
universality conditions, characterizing the ``critical'' regime whence the traveling 
waves are formed, the initial conditions should be sharp enough in impact 
parameter, $i.e.$ $U(t=t_0,r\gg r_0)< e^{-r}$. This condition is
fulfilled by  considering an initial  Gaussian form $e^{-r^2/4B}$ in 
impact-parameter\footnote{It  
corresponds  by Fourier transform to an exponential $e^{-Bk^2_T}, i.e.$ a simple 
diffraction peak in transfer momentum for the 
elastic 
cross-section.}. We will keep this point-of-view in the 
following by considering $e.g.$ a supercritical bare pomeron equipped with a Gaussian form in impact-parameter.

Indeed, in  the critical regime of high rapidity, the  ``very 
forward'' region is driven by the initial condition, while the  ``leading 
edge''
one develops a ``universal'' behavior \cite{wave1} where three terms of the asymptotic 
expansion of the amplitude do not depend either on the details of the  initial 
condition or on the nonlinear damping term. The  ``wave interior''  posseses an 
exact scaling property (see further)  while the ``saturation'' region  depends on the nonlinear term. So, for completion, we will also give some hint on this region in the particular case of the initial RFT  $(\ka=1)$ with a triple Pomeron coupling.
 
 In fact the main universal property of the traveling wave solutions in the 
deterministic case is
 the {\it scaling} property, namely
 \eq
 U(t,r)\ \equiv\ U(r-r_s(t))\ ,\la{scalingfond}
 \eqx
 where the ``time'' dependent radius $r_s(t)$ plays the role of the ``saturation 
scale'' in QCD \cite{munier}.

\subsection{Universal ``Leading edge'' region}

 Let us now derive the traveling wave 
properties in the radial case. For  the leading edge, following a 
similar procedure for the 1-dimensional problem \cite{brunet,beuf},
 one introduces in \eqref{radial1} an ansatz
\begin{equation}
 U\left(t,s=r\!-\!v_c t\right)\  \propto \ 
\exp{\left[-\g_c(s+c(t))\right]}\ 
t^\al\  G\left(\f{s+c(t)}{t^\al}\right) , 
 \la{ansatz}\end{equation}
where $v_c $ ($resp.\ \g_c$) is the critical wave velocity ($resp.$ critical 
slope) of the traveling wave front and $c(t)$ describes the sub-asymptotic 
correction to the velocity  $v(t) \equiv v_c +\partial c(t)/\partial t$. 
This 
ansatz 
 describes the ``velocity blocking'' due to the critical mechanism. 
The 
 point is that, being situated in the forward region where the non-linear
terms in  \eqref{radial1} may be neglected,
 the form of the ansatz can be deduced 
from the linear part of the deterministic  equation \eqref{radial1}. The only effect of the non-linearity is to
ensure the ``velocity blocking'' by the compromise between the fast moving very-forward
regime and the damping due to the non-linear unitarity bound (see, $e.g.$ \cite{munier}).

Inserting  the {ansatz} in the equation \eqref{radial1} and neglecting the 
small 
 contribution from the nonlinear term to the leading edge, we can verify 
the 
equation for the dominant terms (successively  in $t^0, t^{-1/2}, t^{-1}$) 
of 
the time expansion, see Appendix {\bf A}. Note that
the condition  $G(z)\to z$ when $z\to 0,$ is required in order  to match with the 
scaling 
region
(called the ``wave interior'' in \cite{wave1}). 

Adapting to the radial case the standard procedure \cite{brunet,beuf}, one finds
\begin{eqnarray}
U(r\!-\!r_s,t) &\sim& \ (r\!-\!r_s)\ \exp{\left\{-\ (r\!-\!r_s)-\ 
\f{(r-r_s)^2}{\!4\ t}\right\}}\ ,
\nonumber \\
r_s &=& v_c t + c(t)=2t -{\bf 2}\log t\ ,
\label{results}
\end{eqnarray}
where  $r_s$ is the average 
time position of the wave front 
(or ``saturation scale'' in the language of QCD \cite{munier}). Note that 
the 
form of the leading-edge front is the same as  
the one obtained in the 1-dimensional problem \cite{brunet,munier} while the 
saturation scale is $r_s= 2t -{\bf 2}\log t$ instead of $r_s= 
2t -{\bf \f 32}\log t,$ due to the contribution of the new term $\f 1r 
\partial_{r} U$
characteristic of the 2-dimensionality of the initial physical picture. The 
saturation scale evolution is 
thus  slower by a logarithmic factor 
${\bf \f 12}\log t.$ This $\bf 1/2$ shift  is  due to the purely geometrical ``curvature contribution'' of the 
2-dimensional problem \cite{derrida} which combines  with the   coefficient  $\bf 3/2$  of the  FKPP solutions. A 
third (and last) universal term in $r_s$ behaving as $t^{-1/2}$ can also be 
derived and will add some new curvature contributions.

The scaling \eqref{scalingfond} is recovered from \eqref{results} in the  region
$\f{(r-r_s)^2}{\!4\ t}\ll 1$ giving rise to the simple expression 
\begin{eqnarray}
U(r\!-\!r_s) \sim \ (r\!-\!r_s)\ \exp{\left[-(r\!-\!r_s)\right]}\ ,
\label{resultsscal}
\end{eqnarray}
where it ensures the transition with the ``wave interior'' domain.

\subsection{Circular wave properties for a triple Pomeron coupling} 
\subsubsection{Deep ``Wave interior'' region}
Taking into account now  the specific quadratic nonlinear term of \eqref{radial1}  (reflecting 
the original
triple Pomeron coupling), we can explore the deep ``wave interior'' regime, 
adapting 
a method \cite{logan} used for the similar problem in the QCD case 
\cite{parametric}. 

Considering a scaling ansatz with an expansion in a small 
parameter $\Delta^{-2}$ to be determined by consistency with the scaling form \eqref{resultsscal}, 
\ba
U\equiv U\left(z\equiv\f{r-\int^{t} dt' v(t')}\Delta\right)=U_0 + 
\Delta^{-2}\ U_2 
+\Delta^{-4}\ U_4+\cdots\ ,
\la{para0}
\ea
one obtains (see appendix {\bf B})
\ba
U=\f1{1+ e^{z}} + \Delta^{-2}\ \f{e^z}{\left(1+e^z\right)^2}\ 
\log{\f{\left(1+e^z\right)^2}{4e^z}}
+{\cal{O}}(\Delta^{-4})\ .
\la{para}
\ea
with
\ba
z\equiv \f {r-\int^{t} dt' v(t')}\Delta \Leftrightarrow \f {r-r_s}{2}\ ,
\la{sol}
\ea
where the  equality  isobtained
 by matching\footnote{The 
matching between \eqref{sol} and the leading edge velocity \eqref{results} 
is 
not exact at sub-leading level, see Appendix {bf B}.} at high enough $t$ with the critical velocity of the 
leading-edge solution,  namely $\Delta=2.$  Note that it is easy to determine higher order terms by a 
system 
of nested linear differential equations. 
\subsubsection{``Saturation'' region}
\la{satur}
The circular traveling waves being concentric around $r=0,$ the saturation region at 
small $r$ is naturally expected to be different from the 1-dimensional one depicted in Fig\ref{2}, where saturation starts from $-\infty.$ This is 
also 
made explicit by the $\partial_r U/r$ term in Eq.\ref{radial1}, which can no 
more be neglected or considered as giving second order effects  as in the previous regions. For describing the 
saturation region, it is useful to go back to the full 2-dimensional form 
\eqref{FKPP2}. It is convenient to introduce the S-Matrix element $S=1-T,$ which 
is expected to be small in the saturation 
region. In those terms the deterministic\footnote{In fact, the 
noise term would not play a big role anyway, since it is expected to have  small effect in  the ``dense medium'' characteristic of the 
saturated phase.} 2-d equation writes
\eq
 \frac d{dt} \ S\left(t,\vec r\right)= \nabla^2_r S - \ S +\ S^2  
%\epsilon\ \sqrt{1-S}\ \nu(t,\vec r)
\ ,
 \la{FKPPS}
\eqx
Taking into account that  $S^2$ is negligible, Eq.\eqref{FKPPS} boils down to a 
linear equation whose
radial solution is easy to obtain if one notices that $e^{-t}S=W$ is a solution 
of the two-dimensional heat equation, namely $\frac d{dt} \ W= \nabla^2_r\ W.$ A 
simple, azimuthal-invariant solution is thus:
\eq
S(t,r)\equiv1-U(t,r)=e^{-t}\ W(t,r)=e^{-t}\ \left(a-b\ e^{-\f{r^2}{4t}}\right)\ ,
\la{radialS}		
\eqx
where the constants  $a,b$  have to be determined by matching with the wave-interior region. This result shows the general feature of a 
time evolution 
towards the black disk limit 
$S=0,$ at large t.
It does that in a non scaling way, since the approach to the black disk is not 
characterized by a single function  $r-r_s(t).$ 

 \section{Circular traveling wave: stochastic case}
\la{stoc}
\subsection{Quantum fluctuations and the Langevin equation}

As known from the seminal studies of Ref.\cite{brunet}, the effect of even very small fluctuations has an 
important impact on the solutions of the sFKPP equation. They may  drastically modify the solutions of the sFKPP equations  compared to the 
deterministic FKPP ones described in the previous section. Indeed, in the standard sFKPP case, one has been 
able to  analyze  
\cite{mumu} that the  small noise contribution has two superimposed effects. In the 1-d case, the typical expansion  parameter  appears to be  not $\eps$ itself but $1/\log\eps,$ that is the inverse  
{\it logarithm}  of the noise 
strength. 
 At first  order (starting  in fact as   $1/\log^2\eps$) the correction has negative sign and  corresponds to
an  effective cut-off on the amplitude as in  \cite{brunet}. At the next order $1/\log^3\eps,$ a positive contribution comes from rare but large fluctuations of the noise.

  In fact 
we shall now show that
 formula \eqref{sol1} for the analysis of the wave interior in the 
deterministic case the  circular traveling waves with noise can be 
analyzed in a similar  way than for the 1d sFKPP 
case. However, some modifications will be due to the radial extension. Indeed, 
 considering the initial Langevin equation 
Eq.\eqref{radial}, and 
the relation between the noise strength and the effective cut-off 
approximation \cite{brunet,mumu,beuf}, we are naturally led to 
 an effective noise strength 
\ba
\zeta(t) = \ {\epsilon\ {\left[ 2\pi r_s(t)\right]^{-1/2}}}\ ,
\la{cut(t)}
\ea
where in \eqref{radial1} we have substituted ${\f 1{2\pi r}}\to {\f 1{2\pi r_s(t)}}$ in 
the expression of 
 the cut-off. 
 Indeed, this approximation can be  justified by the accompanying factor $U(1-U)\sim 0$ outside $r\sim r_s.$  We see that for the radial case, the effective noise strength depends itself on the saturation scale and thus will possess a rapidity 
dependence. 

In fact the geometrical meaning of the noise strength \eqref{cut(t)} is quite transparent. It takes into account the fluctuations at the periphery of the expanding disk in an azimuthally symmetric way. As an important consequence, the rapidity dependence of the effective noise will play an important physical role role, 
both at  weak and  strong noise regimes, as discussed now.

\subsection{Stochastic traveling waves : weak noise}
\la{secsmall}
Let us solve the 
weak 
noise regime of \eqref{radial}. Noting that choosing the  variable 
 $z=r-\int^{t} dt' v(t')=r- 
2 t+\f1{2}\log(4t-1)$
allows to take into account the radial term in the deterministic part of \eqref{radial}and to  match the 2d radial case with the standard 1d case (up to the  modification \eqref{cut(t)} of the noise). 
 Hence, taking into 
account the parallel properties  of the radial equation
 with the 1d,   it 
is justified 
to export the detailed results obtained for  the 1d sFKPP equation \cite{mumu}. However an important modification of the discussion  
for the radial configuration will appear due to the  time-dependent noise strength 
\eqref{cut(t)}. 

 The 
detailed effect 
of fluctuations has been derived \cite{mumu} and leads to 
the 
following results:
\begin{eqnarray}
U(r,t) &\sim& \ \log{\left(\f {4\pi r_s}{\epsilon^2}\right)}\ 
\sin{\left\{\f{\pi (r-r_s)}{\log{\left(\f {4\pi r_s }{\eps^2}\right)}}\right\}}\ 
e^{-(r-r_s)}\ ,
\nonumber \\
r_s &=& t\left\{2-\f{\pi^2}{\log^2{\left(\f {4\pi r_s 
}{\eps^2}\right)}}+6\pi^2\f{{\log\log\left(\f {4\pi r_s 
}{\eps^2}\right)}}{{\log^3{\left(\f {2\pi 
r_s}{\epsilon^2}\right)}}}+\cdots\right\}\ 
.
\label{resultsstoch}
\end{eqnarray}
The result for 
the stochastic
 average over the amplitude is given \cite{iancudis,gregory} within some approximation \cite{complete}
 by
%----------------------------------------------------------------, 

\begin{equation}
\label{frontU}
\langle U(r,t) \rangle \propto  \erfc\left(\frac{r-r_s}{D\sqrt{t}}\right)
+ \exp\left( \f{D^2{t}}4-(r-r_s)\right)\left[
2 - \erfc\left(\frac{r-r_s}{D\sqrt{t}}-\frac{ D\sqrt{t}}{2} 
\right)
\right]\ ,
\end{equation}
%-------------
where $\erfc(x)$ is the complementary error function and 
\begin{eqnarray} 
 D=\f {2\pi^2}{3\log^3{\left(\f {4\pi r_s}{\epsilon^2}\right)}}
\label{dispers}
\end{eqnarray}
is the stochastic dispersion of the front.
A complete description of the stochastic front, re-summing over all higher moments of the amplitude at weak noise, can be found in 
Ref.\cite{complete}. 

On a more 
general ground, as shown in \cite{iancudis} and suggested  by  numerical 
simulations for the QCD case \cite{gregory}, one predicts a structure of ``diffusive scaling'', namely
\begin{eqnarray} 
U(r,t) \sim U\left\{\f{r-r_s}{2D \sqrt t}\right\}\ ,
\label{disperseff}
\end{eqnarray}
where the parameter $D$ is a characteristic diffusion coefficient, which may differ from the asymptotic \eqref{dispers}. All in all, the solution of the stochastic equation 
can be understood as a dispersive distribution of event-by-event traveling 
waves with dispersion $D.$ The random superposition of traveling 
waves transforms the ``geometric scaling'', valid for each of them into a 
``diffusive scaling'' property \eqref{disperseff} for the average defining 
the final  solution for the amplitude. However, following the numerical 
studies in the framework of QCD \cite{gregory},  diffusive scaling may 
require some evolution time to develop a sizable diffusion coefficient and thus 
may not  be distinguished from  ``geometric scaling'' at physical rapidities.
\begin{figure}[ht]
\begin{center}
\mbox{\epsfig{figure=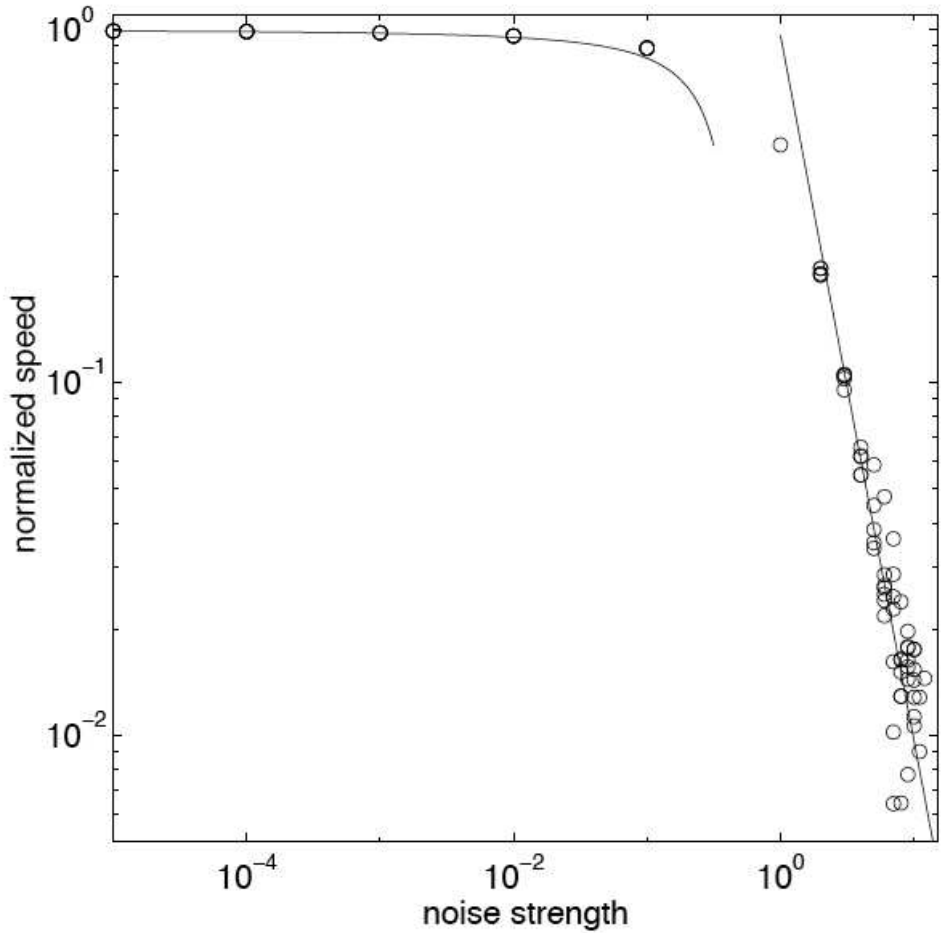,width=15cm,angle=0}}
%\mbox{\epsfig{figure=I_mu.EPS,width=8.6cm,angle=0}} \caption{.}
\la{nois}
\caption{{\it Average wave speed as a function of the noise for the
sFKPP equation.}
{\it Vertical axis}: $v/v_c$ is the average traveling wave speed 
normalized to the speed $v_c=2$ of the deterministic FKPP 
equation; {\it Horizontal axis}: dimensionless noise strength; 
{\it Dots}: numerical results; {\it Left line}:  weak noise analytic prediction; {\it Right 
line}:  strong noise prediction. One 
observes (and may derive \cite{maximal}) a maximal speed around a noise strength 
of order 10. The figure is  from ref.\cite{mueller}.}
\end{center}
\end{figure}

\subsection{Stochastic traveling waves : strong noise }
\la{secstrong}
When 
the noise strength is tuned to increase, one observes a strong decrease of the 
average wave velocity, with a neat change of regime in the 
vicinity of a (normalized) noise strength of order one, see Fig.2. Following 
Ref.\cite{strong}, the overall properties of the 
strong noise regime  are as follows

The solution of equation  \eqref{radial1} is a stochastic average of  
traveling waves   at an average speed 
\eq
v=\frac{2}{\zeta^2} \sim \ \f { 4\pi r_s(t)}{\epsilon^2} 
\ ,
\la{vnoise}
\eqx
where $\zeta$ is the normalized noise strength defined in \eqref{cut(t)}.
Hence the saturation scale $r_s$ follows from the equation
\eq
v\equiv \f{dr_s}{dt}= \ \f { 4\pi r_s(t)}{\epsilon^2} \Rightarrow\  
r_s(t)\propto e^{4\pi t/\eps^2}\ ,
\la{rsnoise}
\eqx
where the rapidity dependence of the radial noise plays the important role. Note that the limiting speed condition anyway requires $v<v_c\equiv 2$ and thus from \eqref{vnoise} ${4\pi r_s(t)}<{\eps^2}.$ Hence  
the exponential behavior of \eqref{rsnoise} must break down before the time evolution reaches  the limit defined by ${2\pi}\ e^{4\pi t/\eps^2}\to {\eps^2}.$ In fact, due to the rapidity decrease of the effective noise $\zeta$ of \eqref{cut(t)}, the strong noise regime 
transforms progressively into the weak noise one, following from right to left the velocity curve depicted in Fig.2.

In the strong noise regime, there exists \cite{strong} an analytic solution for the average solution of 
the evolution equation \eqref{radial1}, namely
\begin{eqnarray}
\label{eq:sol}
\cor{U(t,r)} & = & \frac{1}{2}\ {\rm 
erfc}\left(\frac{r-r_s(t)}{2\sqrt{t}}\right)\nn
& = &\frac{1}{2\sqrt{\pi t}}\int_{-\infty}^\infty dr\,\theta(\tr-r)
                         \exp{-\frac {(\tr-r_s(t))^2}{4t}}\ ,
\end{eqnarray}
where ${\rm erfc}(x)$ is the complementary error function.

This result confirms the decrease of the velocity with increasing noise 
strength. It contrasts with the speed obtained in the weak noise 
limit by perturbative analysis around the F-KPP speed $\simeq 2-\pi^2 
|\log(2\zeta^2)|^{-2}.$
The expression \eqref{eq:sol} shows that the amplitude could be obtained from a 
superposition of step functions around $r_s=v\ t$ with 
a Gaussian form of width $\sqrt{2Dt}$. The interesting point here lies in the 
dispersion coefficient: in the weak-noise analysis, it 
behaves like $|\log(2\zeta^2)|^{-3}$. We have thus shown that the dispersion goes to a 
constant value $2$ when the noise becomes strong.

\section{The Pomeron as a circular traveling wave}
\la{modifs}

Let us investigate the implications of the traveling 
wave 
properties
% summarized in the points i-iii)
 on the soft Pomeron, within the framework that the soft interaction 
dynamics at high energies be governed by a 
``supercritical'' bare Pomeron input. As we have shown in the previous theoretical 
sections,  the circular traveling wave solutions are expected to appear 
due to the combined effect of the high energy evolution and of unitarity, which we will assume to be saturated\footnote{As we have seen in section III, the saturation limit at $b=0$ is not  $T\equiv 1$ at finite $Y$, contrary to the 1-d problem, see Fig.\ref{2}. It reaches $T=1$ when $Y \to \infty.$}. Our analysis will be concerning the asymptotic regime of the circular traveling waves, leaving for further study the transition to this regime.

 We have seen that the evolution towards the saturation limit may
depend on the non-linear terms, see Eq.\eqref{FKPPS}. Those terms
may be physically more complicated than the single quadratic term of 
Eq.\eqref{langevin}.  Hence we will focus on the ``universal predictions'', $e.g.$ those which do not depend on the initial conditions and/or the structure of the non-linear damping. On the contrary, the parameter $\ka,$ which is not fixed by the unitarity constraints, is the relevant parameter, playing an essential role in the Pomeron properties.

\subsection {``Phase diagram'' as a function of noise}
\la{vs}
The first step is to  discuss  which evolution regime we have as a function of $\ka.$ For this sake, the noise strength \eqref{cut(t)} can be conveniently written, restoring the Pomeron variables \eqref{redef}
\ba
\zeta^2 = \f {\eps^2} {2\pi r_s} \equiv  \f \ka{2\pi} \f {\sqrt{\alp \mu}}{b_s} \ .
\la{cut(Y)}
\ea
It is important to note the following feature of the noise strength directly related to the (2+1)-d property of the RFT problem: it decreases together with the expansion of the impact-parameter disk and thus evolves towards weaker noise. However, this decrease, being governed by  the evolution of the disk may be slow. 

The basic relation we will get  comes from the structure of the wave speed reproduced in Fig.2. It is obtained for the 1-d case, but it happens to be indicative also for the radial case, whose universal properties are essentially similar, as we shall see. In Fig.2, one may distinguish how the three different regimes we have analyzed in the previous sections, namely the zero, weak and strong noise respectively, can be identified on the plot where the $\ka$-dependent normalized speed $v_\ka/v_c$ is displayed as a function of the normalized noise strength $\zeta$. With our notations and using \eqref{cut(Y)}, we write by straightforward relations
\ba
\f {v_\ka} {v_c} \equiv \f 1{v_c}\ \f {dr_s}{dt}= \f 1{2\sqrt{\alp\mu}}\ \f {db_s}{dY}
= \f \ka {4\pi}\ \f 1{\zeta^{2}}\ \f {2db_s}{b_s dY} \ ,
\la{vka}
\ea
where we have denoted $v_\ka \equiv \f {dr_s}{dt}$ the actual wave front  velocity and
$v_c =2,$ the deterministic critical speed. Note that we have made use of \eqref{cut(Y)} to  
substitute the bare Pomeron parameters $\sqrt{\mu \alp}$ by its expression in terms of the normalized noise. Our final expression thus  writes
\ba
\f {v_\ka} {v_c} =  \left[\f {\ka\dl} {4\pi}\right]\  {\zeta^{-2}}\ ,
\la{final}
\ea
where $\dl \equiv 2 {db_s}/{b_s dY} = d\log {\cal A}/{dY},$ where  ${\cal A}=\pi b^2_s$ is the area of the effective impact-parameter  disk for the collision.
The obtained expression shows directly how the noise strength $\ka$ parametrizes the normalized-speed $vs.$ normalized-noise relation depicted in Fig.2. It allows one to relate the ``phase diagram'' defining the different regimes of the radial sFKPP equation to a  physical soft Pomeron feature, namely the exponent $\dl$ of the expanding disk area ${\cal A}$.

When  interpreting Fig.2, one may distinguish the different regimes as follows using relation \eqref{vka}:

\bit
\item
The weak noise regime:
\ba
\!\!\!\!\!\!\!\!\!\!\!\!\zeta \le {10^{-1}} \quad \quad .9\le \f {v_\ka} {v_c}\le 1. \quad \quad\quad  \ \f \ka {4\pi} \le \f  {10^{-2}}\dl \quad \quad 2\sqrt{\mu \alp} \lesssim\ { \f 12} {\dl b_s}  = \ \f {db_s}{dY}\ .
\la{wn}
\ea
\item
 The strong noise regime: 
\ba
\ \quad \zeta \gtrsim 1.4 \quad \quad \quad  \zeta^{-2}=\f {v_\ka} {v_c}\ \lesssim .5 \quad \quad  \ \ \ \f \ka {4\pi} 
= \f 1 \dl \quad \quad \quad \ 2{\sqrt{\mu \alp}} \sim  \f {\zeta^2}2\ {\dl b_s}= {\zeta^2}\f {db_s}{dY}\gtrsim 2\f {db_s}{dY}\ , 
\la{sn}
\ea
where we made use of the exact relation at strong noise \eqref{vnoise}. 
To complete the picture, one adds
\item
The zero noise regime: 
\ba
\!\!\!\!\!\!\!\!\!\!\!\!\!\!\!\!\zeta \lll 1 \quad \quad  \quad \quad \f {v_\ka} {v_c}\sim 1  \quad \quad  \quad\quad\quad \f \ka {4\pi} \sim \ 0 \quad \quad  \quad {2\sqrt{\mu \alp}} \sim\ \f {1}2\ {\dl b_s}=\ \f {db_s}{dY}\ .
\la{zn}
\ea
\item
The middle noise regime: 
\ba
\!\!\!\!\!\!\!\!\!\!\!\!\!\!\!\!\!\!\!\!\!\!\!\!.1\le\zeta \le 1.5 \quad \quad  .2\le\ 
\f {v_\ka} {v_c}\ \le .9 \quad \quad \f  {10^{-2}}\dl\le \f \ka {4\pi} \le \f 1 \dl\quad \quad  \quad \  \  \f {db_s}{dY} \le{2\sqrt{\mu \alp}} \le\  2\f {db_s}{dY}\ .
\la{mn}
\ea
\eit

We see from relations (\ref{wn}-\ref{mn}) that the parameter $\ka$ plays the role of the order parameter of the RFT. Once given  the physical observable  $\dl$, one knows the phase ($cf.$ evolution regime) of the system from the determination of $\ka.$ In particular for the original RFT action \eqref{action0}, the phase is completely specified by $\dl.$ Note also that the ``bare'' parameter  $2\sqrt{\alp\mu}$ corresponding to the maximal critical speed of the disk relates to $b_s\dl/2\equiv \f {db_s}{dY}$ which is  a
``dressed'' parameter in terms of a field theory.

The question  of determining the soft Pomeron properties thus boils down to the determination of $\dl .$ We 
postpone a detaileed phenomenological study  for the  future\footnote{A first qualitative phenomenological exploration is discussed in Appendix {\bf D}.},
 but it is not too difficult to evaluate the order of magnitude of  $\dl .$ Indeed, if the black disk limit would
 have been nearly reached, one would expect a geometrical cross-section $\sig_{tot} \propto {\cal A}= \pi b^2_s$
 and thus  $\dl\sim d\log \sig_{tot}/dY\sim .08$ where the last number is the well-known popular determination \cite{donach}. This value for $\dl$ should be considered as a maximum at present energies, since the black disk limit seems not to be fully reached (see, $e.g.$, \cite{models}, where one obtains smaller values of order $\dl\sim (1-3)10^{-2}$). as an example in appendix {\bf C} we show the phase diagram characteristics when choosing  the conservative values of $\dl \sim 10^{-2}$ and $b_s \sim 1\ {\rm fermi = 5\ GeV}^{-1}.$  

In any case, one interesting remark is that for the whole range $\dl \in [1-8]10^{-2},$ the original RFT with $\ka=1$ seats within the limit of the weak noise region. By comparison, the  order parameter $\ka$ takes a factor 100 in the interval between the weak and the strong noise regimes\footnote{It has been shown using field theory arguments\cite{maximal} that a maximal noise exists at $\zeta=8\pi$ beyond which the traveling waves stop and the system does no more ``percolate'', $i.e.$ the disk stops expanding with rapidity.}. Such high values of $\ka$ are not {it a priori} forbidden, even if far from the original RFT action with a single triple Pomeron coupling. We will discuss in conclusion a possible QCD interpretation of these large values of $\ka$.

\subsection{Properties of the ``wave front''}
\la{front}
\subsubsection{An expanding impact-parameter disk}
\la{expand}
Having now identified the phase diagram of the RFT solutions, we are able to 
discuss  the characteristic features of the soft Pomeron as a circular traveling way by
recasting the results of section II ($resp.$ III) for the deterministic ($resp.$ 
stochastic) traveling waves in terms of  the physical variables through the relations \eqref{redef}.

As a general result, valid in all cases, we find that   the front of the 
traveling wave is situated around the impact-parameter value 
\eq
b_s(Y)=\sqrt{\alp/\mu}\ r_s(t=\mu\ Y)\ .
\la{frontY}
\eqx
The expanding  impact-parameter disk   is thus related  to the increasing function $r_s(t).$ This
is the analogue of  the rapidity-dependent ``saturation scale'' \cite{GW} discussed also in the framework of QCD 
traveling waves \cite{munier}. Let us examine the equivalent ``saturation scale'' of the supercritical Pomeron. As we 
have seen previously, it depends on the phase diagram. Following 
Eqs.\eqref{results}, \eqref{resultsstoch} and \eqref{vnoise} respectively, we 
obtain
\eq
b_s(Y)- b_s(Y_0)&=&2 \sqrt{\alp\mu}\ \left[(Y\!-Y_0)-\f 1\mu\ {\log \f Y{Y_0}} + \cdots\right],\ {\rm for\ zero\ 
noise}\nn
b_s(Y)- b_s(Y_0) &=& 2\sqrt{\alp\mu}\ (Y\!-Y_0)\ \left[1-\f{\pi^2}{2\log^2\left(2\zeta^{-2}\right)}+\f {3\pi^2\log\log\left(2\zeta^{-2}\right)}{\log^3\left(2\zeta^{-2}\right)}+\cdots\right],\ {\rm for\  weak\  noise}
\nn
b_s(Y)&=& b_s(Y_0) \quad \exp{\left[\f{4\pi}{\kappa}\ (Y\!-Y_0)\right]} ,\ {\rm for\  strong\  noise}.
\label{diskall} 
\eqx
The middle-noise regime does not posses an analytic expression but its  numerical implementation is possible and shown (in the reduced variables) on Fig.2. The dots $(\cdots)$ stand for sub-leading and/or non-universal higher order terms. Note that the constant terms implied by the initial condition at $Y\!=Y_0$ are also naturally not constrained by universality. The validity of relations \eqref{diskall} thus require a large enough interval $Y\!-Y_0.$
For all cases
one finds from  \eqref{diskall} that the disk expands  
with
rapidity. For the ``zero noise'' and  ``weak noise'' cases the asymptotic 
velocity is $2 \sqrt{\alp\mu},$ which is the critical velocity, and thus driven 
by the deterministic equation. The 
situation is different at ``strong noise'' where the solution is still moving 
linearly with rapidity $Y$ but with a velocity governed by the order parameter
$\ka$. 

For the ``zero noise '' and  ``weak noise'' cases, we have thus identified new universal 
terms\footnote{Note that a third 
universal term, behaving as $1/{\sqrt{Y}}$ is expected to exist in the 
asymptotic expansion of the deterministic case from 1-d studies 
\cite{wave1,munier}. We leave its 
determination in the present 2-d case for further study.} not depending on the initial conditions nor on the specific 
non-linear damping terms. The existence of an universal rapidity expansion
  due to the supercritical Pomeron is, to 
our knowledge, a new result allowed by the Langevin formulation of the RFT and its traveling wave solutions. Moreover they appear to be quite different depending on non trivial phase diagram: in the 
deterministic case, the first (negative) correction to the radius $b_s\propto Y$ behaves like $\log Y$, while it is of order $Y\log^{-2}Y$ 
for  ``weak noise'', and thus {\it a priori} quite more important than in the deterministic case. 

For ``strong noise'', the obtained exponential behavior would  not lead ultimately to a violation of the
Froissart bound since the  rise  is tamed by  the boundary of the strong noise regime.  From \eqref{cut(Y)} and the relations \eqref{sn}, one has
\ba
\la{limit}
\zeta^{-2} = \f {2\pi b_s}{\ka\sqrt{\alp\mu}}= v_\ka/v_c\lesssim .5\quad \Rightarrow\quad b_s\  \lesssim \f {\ka}{4\pi}\ \sqrt{\alp\mu}\ . 
\ea
In fact this limit on the impact-parameter disk reflects the $Y$-dependence of the noise strength and thus the evolution from strong noise towards weak noise through an intermediate middle-noise regime. Interestingly enough, this would mean for the radius (and thus for the cross-section near the black disk limit) a gradual transition from an exponential towards a squared logarithmic behavior in  rapidity and thus an asymptotic restoration of the Froissart bound.

\subsubsection{Impact-parameter Scaling}
\la{Scal}
The scattering amplitude is related to the front profile of the dominant 
asymptotic 
traveling wave solution through
\eq
T(Y,b)=U\left(t=\mu\ Y,\ r=\sqrt{\f \mu \alp}\ b\right)\ ,
\la{frontYb}
\eqx
at least in the region where universal results apply. $U(t,r)$ being given 
by formulas
\eqref{scalingfond}, \eqref{disperseff} and \eqref{eq:sol} respectively, one obtains
\eq
T(Y,b) &\sim&  T\left\{b-b_s(Y)\right\},\ {\rm for\ zero\ 
noise}\nn\\
T(Y,b) &\sim& T\left\{\f{b-b_s(Y)}{D_w(Y)\sqrt{\alp (Y\!-Y_1)}}\right\},\ {\rm for\  weak\  
noise}
\nn
T(Y,b) &\sim& \ T\left\{\frac{b-b_s(Y)}{D_s\sqrt{ 
\alp(Y\!-Y_1)}}\right\},\ {\rm for\  strong\  
noise}\ ,
\label{profileall} 
\eqx
where the values of the front scales $b_s(Y)$ have to be chosen from \eqref{diskall} for each corresponding individual  regime. $D_w$ and $D_s$ are the dispersion parameters for  weak and strong noise respectively.
We have introduced a rapidity scale $Y_1$ which denotes the effective rapidity where the dispersion of the noisy traveling waves becomes  sizable. Indeed, numerical simulations for the QCD case \cite{gregory} show that such a threshold do exist. Below that threshold the scaling is similar to the zero noise case.
From \eqref{dispers} and  \eqref{eq:sol}, one finds
\ba
\la{disp}
D_w=\f{2\pi^2}{3\log^3\left[2\zeta^{-2}\right]}=\f{2\pi^2}{3\log^3\left[4\pi b_s/\ka\sqrt{\alp\mu}\right]}\ ;\quad\quad D_s\equiv 2\ .
\ea
It is clear that the dispersion parameter $D_w\propto 1/\log^3Y$ may be quite small and thus one would then recover the same scaling as the zero-noise regime, but with the different weak-noise evolution  $b_s(Y)$. Moreover the dispersion is proportional to $\sqrt{}\alp$  which may be small
even for a sizable value of $\sqrt{\alp\mu}$ for larger $\mu.$

\subsubsection{Front profile}
\la{profile}In the regions where there exists a universal form of the wave front profile, 
$i.e.$
within and forward to the wave front (the previously called ``wave interior'' and ``leading edge'' regions), one finds
from formulas 
\eqref{scalingfond}, \eqref{frontU} and \eqref{eq:sol} respectively,
\eq
T(Y,b) &\sim&   (b-b_s)\ 
\exp{\left\{-\sqrt{\f \alp \mu}\ (b-b_s)-\f {\left(b-b_s\right)^2}{4\alp (Y\!-Y_0)}\right\}},\ 
{\rm for\ zero\ noise}\nn
%\eqx
%\end{document}
T(Y,b) &\sim&  \erfc\left\{\frac{b-b_s}{D_w\sqrt{\alp(Y\!-Y_1)}}\right\}+ \exp{\left(\frac{\mu\ D_w^2{(Y\!-Y_1)}}{4}-\sqrt{\mu/\alp\ (b-b_s)}\right)}\ \times\nn &{}&\ \  \times \ \left\{2 - \erfc\left(\frac{b-b_s}{D_w\sqrt{\alp(Y\!-Y_1)}}-\frac{D_w\sqrt{\mu(Y\!-Y_1)}}{2} \right)\right\},\ {\rm for\  weak\  noise}
\nn
 T(Y,b) &\sim& \erfc\left\{\frac{b-b_s}{2\sqrt{\alp
(Y\!-Y_1)}}\right\},\ {\rm for\  strong\  
noise}\ ,
\label{profileall1} 
\eqx
where $D_w,D_s$ were given in \eqref{disp}.

Some comments are in order about the front profiles.
We may note that $T(b,Y)$  for  the zero noise case\footnote{It is also true for the noisy
traveling wave solution \eqref{resultsstoch} for weak noise.}
goes to  zero with $b-b_s,$ corresponding to the 
way 
how  saturation is imposed as an ``absorbing  condition'' \cite{absorb} 
to  the leading-edge approximations of the traveling waves' tails. 
Around and below $b=b_s$  the solutions  get corrections to the spurious zero.
However, these corrections (see, $e.g.$ \eqref{para}) are more dependent on the specific form of the
 nonlinear
terms of the equation.

Note also that the diffusive scaling \eqref{disperseff} is exact for the strong 
noise case and with exact dispersion 
parameter $D_s\equiv 2.$ 
It comes from an exact solution of the  statistical mechanic 
picture \cite{strong}.
Indeed, the strong noise regime can be interpreted as the stochastic superposition of 
traveling waves of the simple form $\theta(b-b_s),$ suggested by Eq.\eqref{eq:sol}. Moreover, at strong noise, the average solution is dominated by the strongest fluctuations, together with all correlators, as shown in \cite{strong}.

\section{Conclusions}
\la{out}
%\subsection{Summary}
We can summarize our results as follows:

{\it i)} Reggeon Field Theory, when the bare Pomeron is ``supercritical'', can be formulated as  a stochastic equation which is in the same ``universality class'' than the  2-dimensional version of the stochastic Fisher and Kolmogorov-Petrovsky-Piscounov equation. In this framework, ``time'' corresponds to a rapidity evolution $Y-Y_0$ and ``space'' to the 2-d impact-parameter $\vec b$ of the hadronic collision. 

{\it ii)} Thanks to the mapping to the 2d-sFKPP equation, one is able to find the asymptotic and certain subasymptotic solutions of the RFT which were beyond  reach of the purely field theoretical methods used in the past. These solutions possess an appealing ``universality '' property, which means that they do not depend either from the initial conditions  or on the peculiar form of the non-linear terms ensuring the unitarity constraint on the elastic amplitude.

{\it iii)} To our knowledge, it is yet the only example of a supercritical Pomeron theory preserving an universal behavior of the amplitude. Usually, the factorisation property of a Pomeron as a Regge pole, on which relies the standard universality arguments (see, $e.g.$ \cite{donach}), is expected to be washed out 
by interactions for a RFT based on a supercritical Pomeron. This ``universality'' property is recovered in a very different way, since it comes from a dynamical mechanism based on the ``critical'' phenomenon associated to the formation of circular traveling waves.

{\it iv)} The ``universality class'' property remains valid  when the splitting and merging Pomeron vertices are of ratio $\ka\ne 1,$ which was the RFT value  based on the unique triple Pomeron coupling. Indeed, $\ka,$ which is a measure of the  strength of the ``Pomeron loop'' contribution, plays the role of the ``order parameter'' in the phase diagram of the RFT. Hence the original RFT (with $\ka=1$) lies in a specific phase of the diagram. A rough but realistic estimate (assuming  RFT to be physically applicable) places the original RFT in the ``weak noise'' regime of sFKPP. However, thhis choice is not dictated by a theoretical constraint. The full phase diagram should allow a model-independent  discussion of this degree of freedom, if compared directly with the phenomenology.

{\it v)} More generally, depending on  the dimensionless noise 
strength $\ka$ the phase diagram is shown to lead to three specific  phases 
corresponding respectively to zero noise ($\ka=0$), weak noise 
($\ka={\cal O}(1)$) and strong noise ($\ka={\cal O}(100)$), (plus an intermediate middle noise one (${\cal O}(1)\lesssim \ka \lesssim {\cal O}(100)$), for which explicit asymptotic regimes of solutions  are obtained in the front and in the tail of the impact-parameter disk. Some other results, valid in the whole range are obtained in the case of triple-Pomeron coupling.

There are intringuing theoretical\footnote{The present paper is not devoted to a phenomenologocal study. However, we have listed in appendix {\bf D} an outlook of  possibly relevant phenomenological remarks.} lessons to be drawn from our results
The striking theoretical feature of the approach to the RFT through the mapping to the 2-dimensional sFKPP equation
is its ability to avoid  the complications    of  an usual field theory formulation  in the case
of a ``supercritical'' bare Pomeron. Indeed, if at first a ``critical'' Pomeron theory for which the renormalization group exists raised some hope (see $e.g.$ \cite{abarbanel}) it  led to unphysical results for hadronic reactions such as total cross-sections behaving as $Y^{1/12}.$ For a ``supercritical Pomeron''
the field theoretical methods, interesting as they may be (see $e.g.$ \cite{parisi}), appeared to be technically complex with  difficulties to conveniently handle the solutions. It thus seems that the  traveling wave methods developped in the present study are well suitable for avoiding the obstacles. It is quite remarkable that, even a domain dominated by very large ``quantum loop'' contributions such as the strong noise phase discussed here, can be handled in a quite economic way.

It is useful to list a series of interesting theoretical subjects which lie beyond the present study. One first problem is to get rid of the approximations made for the derivation of the solutions, the main one being to have replaced in the original equation \eqref{radial} the $r$-dependent coefficients by their value on the wave front $r_s.$ It would improve the analysis to solve, even numerically, the original equation to check the validity of the  approximation.

The question of the azimuthal dependence of the noise is perhaps challenging and thus interesting. Indeed, It could {\it a priori} be possible that azimuthal symmetry be recovered $after$ averaging over the noise. However, it is known from the study of a plane front \cite{kessler} that instabilities may be created by the fluctuations beyond some threshold. It is reasonable to expect that the unitarity constraint should cut-off such inhomogeneities, leaving the purely radial solution valid. For non azimuthally symmetric variables, this is not so obvious and studies $e.g.$ related to the ``supercritical Pomeron'' in diffraction dissociation or for particle production could lead to some interesting problems, such as the noisy structure of the diffraction disk \cite{parisi2}.

Finally, it would be natural and interesting to address the question of the relation of our results with QCD. {\it A priori}, there is a long way to go from an effective and thus ``macroscopic'' theory of Pomeron interactions to a ``microscopic'' point-of view based on quark and gluon interactions. Perhaps a tentative  approach would be to notice that the RFT can be considered using the ``hard'' Pomeron as an input and thus depending of the value of a QCD coupling constant $\al_S$. Since, finally, the only order parameter we have is $\ka,$ this would mean that this parameter should be considered depending on $\al_S.$ It is interesting to note that the strong noise regime leads to the rapidity dependence $b_s \propto e^{4\pi/\ka}.$ A ``hard'' pomeron behaviour which value of intercept is proportional to $\al_S$ would lead to choose $\ka \propto 1/\al_S.$ Hence a ``perturbative'' property for QCD would be in relation with a highly quantum regime (large Pomeron loops strength $\ka$) in terms of RFT. By contrast a quasi-classical regime of RFT (zero or weak noise, small $\ka$) would be associated to a large effective coupling constant of order $\al_S\sim 1/\ka.$ This speculative but intringuing ``duality property'' deserves certainly some interest.
\section*{{\bf{Acknowledgments:}}}
Fruitful discussions with Andrzej Bialas,
Guillaume Beuf, Bernard Derrida, Cyrille Marquet, Emmanuel Saridakis and  comments from Edmond Iancu,  Fran\c cois G\'elis and Maciek Nowak are acknowledged. We thank Alan Martin and Genya Levin for a useful information on their works. The present work has been completed at the Jagellonian University in Cracow, with the support of the VI Program of European Union ``Marie Curie transfer of knowledge'', project: Correlations in Complex Systems ``COCOS'' MTKD-CT-2004-517186.
\eject
\section*{Appendix A: Derivation of the radial ``leading edge''}
\la{generalA}

In the following we shall restrict our analysis  to radial amplitudes, $i.e.$ 
 depending  only on the radial coordinate $r=|\vec r|.$ One starts with 
\begin{equation}
 \frac d{dt} \ U\left(t,r\right)=  \partial_{rr} \ U +\f 1r\ \partial_{r} \ U +\ 
U -\ U^2\ . 
 \la{radialA}\end{equation}

Let us introduce, following a similar procedure for the 1-dimensional problem 
\cite{brunet,beuf}, the ansatz
\begin{equation}
 U\left(t,s=r\!-\!v_ct\right)\  \propto \ \exp{\left[-\g_c(s+c(t))\right]}\ 
t^\al\  G\left(\f{s+c(t)}{t^\al}\right)\ , 
 \la{ansatzA}
 \end{equation}
where $v_c$ ($resp.\ \g_c$) are the critical wave velocity ($resp.$ critical 
slope) of the traveling wave front. This ansatz is for describing the 
universal behavior of the wave in the {\it leading edge} region forward to the 
front \cite{wave1}. 

Inserting \eqref{ansatzA} in the equation \eqref{radialA} and neglecting the 
small  contribution from the nonlinear term to the leading edge, we can verify 
the equation for the dominant terms of the time expansion. the different terms 
give:
\begin{eqnarray}
\f 1{U}\ \f{dU}{dt}&=& \left(\g_c-t^{-\al} \f{G'}{G}\right)v_c-\g_c c'(t)+\f \al 
t+\al t^{-\al-1}[s+c(t)]\f{G'}{G}+...
\nonumber \\
-\f 1{U} \f{\partial_s U}{r}&=& \left(\g_c-t^{-\al} 
\f{G'}{G}\right)\left[\f1{s+v_ct}\right] = \left(\g_c-t^{-\al} 
\f{G'}{G}\right)\left[\f1{v_ct}\right]+...
\nonumber \\
\f 1{U} {\partial_{ss} U}&=& \left(\g_c-t^{-\al} \f{G'}{G}\right)^2 
+t^{-2\al}\left[ \f{G''}{G}-\f{G''^2}{G^2}\right]+...\ ,
\label{eqbasicvar00}\end{eqnarray}
where the dots $(...)$ indicate irrelevant sub-dominant contributions.

Order by order in the late time expansion we get the following relations
\begin{eqnarray}
\g_c v_c=(\g_c)^2 +1\quad &\Rightarrow&{\rm (order}\ t^0)
\nonumber \\
v_c=2(\g_c)\quad&\Rightarrow&({\rm order}\ t^{-\al})
\nonumber \\
0=\al \g_c\beta_c-\al -\f {\g_c}{v_c}+\al 
z\f{G'}{G}+\f{G''}{G}\quad&\Rightarrow& ({\rm order}\  t^{-2\al})\ 
,
\label{relationsA}
\end{eqnarray}
with $z\equiv (s/t^{-\al})$ fixed and finite and $c(t)\equiv \beta_c\log t.$
Hence we get the values of the critical parameters $\al=1/2;\g_c=1;v_c=2.$
The last equation of \eqref{relationsA} now reads
\begin{equation}
0=(\beta ^*-1)\ G(z)+\f z2\ G'(z)+G''(z)\ .
\la{differentialA}
\end{equation}
The condition  $G(z)\to z$ when $z\to 0,$ necessary to match with the scaling 
region
(called the ``wave interior'' in \cite{wave1}), leads to  $\beta_c=5/2$ and 
$G(z)\propto z\ e^{-z^2/4}$. One thus finally gets
\begin{eqnarray}
U^{l.e.}(r\!-\!r_s,t) &\sim& \ \exp{\left[-(r-r_s)\right]}\  
(r-r_s)\ \exp{\left(-\f 14\ \left[\f{r-r_s}{t^{1/2}}\right]^2\right)}\ ,
\nonumber \\
r_s &=& 2t -2\log t
\label{resultsA}
\end{eqnarray}
where ${l.e.}$ stands for {\it leading edge} and $r_s$ is the average 
moving position of the wave front 
(or ``saturation scale'' in the language of QCD \cite{munier}). Note that the 
form of the front is identical to 
the one obtained in the 1-dimensional problem \cite{brunet,munier} but the 
saturation scale is $r_s= s-c(t) + 1/2 =2t -{\bf 2}\log t$ (instead of $r_s= 
s-c(t) + 1/2 =2t -{\bf \f 32}\log t,$) and thus slower by a logarithmic factor 
${\bf \f 12}\log t.$ This shift can be interpreted (and checked) \cite{derrida} 
as resulting from the superposition of the ``curvature contribution'' of the 2-dimensional 
problem with the dynamical slowing down of the 1d FKPP solutions.
\eject
\section*{Appendix B: Derivation of the radial ``wave interior''}
\la{generalB}

Let us introduce the formula \eqref{para0} into the equation \eqref{radial1}
and expand in powers of $\Delta^{-2},$ one gets
\eq
0&=&\left[\f {v(t)}\Delta +\f 1{\Delta\int^{t} dt' v(t')}\right]U'_0 + 
U_0-U_0^2\nn
0&=&\left[\f {v(t)}\Delta\right]U'_1 + U''_0+\left(1-2U_0\right)U_1
\ ,
\la{a1}
\eqx
where, one has replaced $\f 1r\ \partial_{r}U\to \f 1{\int^{t} dt' v(t')}\ 
\partial_{r}U$ in \eqref{radial1} since the difference is $ {\cal O}(1)\ll 
\Delta t$ in the wave interior.

In fact, we choose $\Delta=cst.$ such that $\left[\f {v(t)}\Delta +\f 
1{\Delta\int^{t} dt' v(t')}\right] =1,$ leading by simple integration to the 
equation 
\ba
r_s\equiv \int^{t} dt' v(t')\equiv \Delta t -\f1\Delta\log\left(\Delta 
r_s\right)\sim \Delta t-\f1\Delta\log\left(\Delta^2t-1\right)
\la{sol1}
\ea 
at large time. Solving the simple nonlinear equation of the first line of 
\eqref{a1} one easily gets
\ba
U_0=\f1{1+ e^{\f{r-r_s}\Delta}}\equiv \f1{1+ e^z}\\ .
\la{a2}
\ea
Knowing the solution for $U_0,$ it is not too difficult to solve the linear 
equation, second line of \eqref{a1}, obtaining with the appropriate boundary 
conditions
\ba
U_2=\f{e^z}{\left(1+e^z\right)^2}\ \log{\f{\left(1+e^z\right)^2}{4e^z}}\ .
\la{apara}
\ea

\section*{Appendix C: RFT phase diagram for $\dl=2\ 10^{-2}$ and $b _s \sim 5\ {\rm GeV}^{-1}$}
\la{apply}

\bit
\item
The zero noise regime: 
\ba
\!\!\!\!\!\!\!\!\!\!\!\!\!\!\!\!\zeta \ll 1 \quad \quad  \quad\f {v_\ka} {v_c}\sim 1 \quad \quad \quad  \quad\  \ka  \ll \ 1 \quad \quad  \quad{\sqrt{\mu \alp}} \sim\ .025\ {\rm GeV}^{-1}\ .
\la{azn}
\ea
\item
The weak noise regime:
\ba
\zeta \le {10^{-1}} \quad \quad .9\le \f {v_\ka} {v_c}\le 1 \quad \quad \quad \  \ka  \le {2\pi}  \quad \quad {\sqrt{\mu \alp}}  \lesssim  .025\ {\rm GeV}^{-1}\ .
\la{awn}
\ea
\item
 The strong noise regime: 
\ba
\ \quad \quad \quad \quad \zeta \gtrsim  1.4 \quad \quad \quad  \zeta^{-2}=\f {v_\ka} {v_c}\ \lesssim .5 \quad \quad  \ \ka 
=200\ \pi \quad \quad \quad \ {\sqrt{\mu \alp}} \sim  \f {\zeta^2}4\ {\dl b_s} \sim .025{\zeta^2}\ \gtrsim .05\ {\rm GeV}^{-1}\ , 
\la{asn}
\ea
\item
The middle noise regime: 
\ba
\!\!\!\!\!\!.1\le\zeta \le 1.5 \quad \quad  \quad .2\le\ 
\f {v_\ka} {v_c}\ \le .9 \quad \quad \quad  \quad\  2\pi \le  \ka  \le  200\ \pi \quad \quad  \quad \ 
.025\ {\rm GeV}^{-1}  \le{\sqrt{\mu \alp}} \le\ .05\ {\rm GeV}^{-1}  \ .
\la{amn}
\ea
\eit

\section*{Appendix D: Phenomenological Remarks}
\la{D}

 {\it a)  Scaling in impact-parameter.}
 The main property of the traveling wave solution  is its scaling structure in impact-parameter. On a phenomenological ground,  
assuming for simplicity a purely  imaginary elastic amplitude, one obtains  
 a scaling property of the elastic amplitude considered as a function of  
impact-parameter and energy. One may write $
 ImT_{el}(Y,b) \sim  T(b-b_s),$
 where
$b_s$ describes an  expanding scattering disk ${\cal A} =\pi b^2_s$ possessing 
universal slowering corrections.
Note that it is the ``soft interaction'' version, in the variables $(Y,b)$, of a the similar ``geometric scaling'' \cite{golec}  of the ``hard interaction''
encountered in  deep-inelastic scattering\footnote{The term ``geometric scaling'' 
has been used long ago \cite{dias} for soft reactions at 
lower energies, but with a radically different formulation, namely $T(b,Y)\sim T(b/R(Y)).$} and involving instead the  $(Y,\log Q^2)$ variables.

As discussed in the paper, we expect  scaling  to stay approximately valid at weak noise, at least before a fully realized  stochastic regime takes place where $
 ImT_{el}(Y,b) \approx T\left\{\f {b-b_s}{D\sqrt{Y}}\right\}.$
However, in this case, some non negligible corrections are expected to appear also in the disk radius, see \eqref{diskall}.

 {\it b) Total cross sections}

The  existence of an expanding disk in 
impact-parameter
given by Eqs.\eqref{diskall} may have  a direct consequence on forward scattering amplitudes, and 
through 
unitarity, on the total cross-section at high energy.
Indeed, taking as an example the ideal geometric relation $\sig_{tot}\propto {\cal A}$ one finds for the zero noise case
in appropriate units and for the dominant terms at asymptotic $Y$ 
%, using \eqref{area}
\begin{equation}
\sigma\to 
 \left\{Y^2-\f{2Y\log 
Y}{\mu}+\cdots\right\}\ .
\la{sigma}
\ea
Eq.\eqref{sigma} saturates the $Y^2$ behavior given by the  Froissart bound, and thus restores unitarity. However, our prediction is the existence of a ``universal'' $Y\log Y$ 
correction term with strength governed by the bare Pomeron intercept $\mu.$ It is directly related  to the correction  term of the 
traveling wave speed \eqref{results}.

Note that the result for the stochastic regime may be  significantly 
different, namely
\eq
\sigma(Y)  &\to&  Y^2\left\{1-\f{2\pi^2}{\log^2{\left(\f {8\pi
Y}\ka\right)}}+\cdots\right\},\ {\rm for\  weak\  noise}
\nn
\sigma(Y) &\to& \ \exp{\f{8\pi\ Y}{\kappa} } ,\ {\rm for\  strong\  noise}\ 
,
\label{sigmastochphys} 
\eqx
The results \eqref{sigmastochphys} call for comments. In the  weak noise regime, it is clear that the 
stochastic corrections depending on the parameter $\ka$ are of order $1/\log^2Y$ and thus significantly more important than the 
deterministic ones  of   order $\log Y/Y,$ see
formula \eqref{sigma}. Hence the Pomeron loop effects are expected, if their coupling $\ka$ in the supercritical Pomeron 
scenario is effective, to have an observable effect.

 At strong 
noise, the behavior of the cross-section is entirely governed by the noise, with its characteristic parameter $\ka.$ It is interesting 
to note that, if the phenomenological soft Pomeron of Ref.\cite{donach} with  intercept $1.08$ is attributed to a strong noise scenario, 
it would correspond to $\ka \sim 100\pi,$ which is  discussed in the previous section \ref{model}. Note that anyway the 
noise strength decreases like $1/\sqrt Y$ and  the strong noise regime will transform into the middle if not the weak noise one after some rapidity evolution. Hence
the apparent violation of the Froissart bound will not be maintained at high enough energy.

{\it c) Modification of the large $b$ behaviour}

It is interesting to study how the traveling wave behavior modifies the transfer momentum dependence of the 
elastic cross-section and thus the diffraction peak. Indeed, as we have seen 
previously, starting with an initial condition which is Gaussian in 
impact-parameter, the asymptotic traveling wave solution drives the 
solution of the evolution equation to a different, ``universal'' form ($e.g.$ 
independent from the initial condition and the precise 
form of the non-linear damping terms in the equation). By Fourier transform, 
this evolution should change the structure of the amplitude 
in momentum transfer, and thus modify the diffraction peak. 

For an example we will start with the expression of the 
wave 
front in the ``leading edge'' domain at zero noise, formula  \eqref{results}. With some 
rearrangement of terms one can write
\begin{equation}
 \la{regge}
T(Y,b) \propto (b-b_s)\ \exp{\mu (Y\!-Y_0)}\  
\exp{-\f {\left[b + 2\sqrt{\f \alp \mu}\ \log(\mu (Y\!-Y_0))\right]^2}{4\alp (Y\!-Y_0)}}\ .
\end{equation}
%\begin{equation}
% \la{xsections}
%\frac{d\sigma}{d|t|} \approx e^{b_s^2t} \sim e^{2t \left\{\alp\mu\ Y^2 - 
%\sqrt{\alp\mu}\ Y\log Y\right\}}
%\end{equation}
In Formula \eqref{regge}, we note that, apart the linear prefactor,  the 
exponential behavior dominant at large $Y$ boils down to the following modification
\begin{equation}
 \la{modif}
\exp\left\{\mu (Y\!-Y_0) -\f{b^2}{4\alp (Y\!-Y_0)}\right\} \to  
\exp\left\{\mu (Y\!-Y_0) -\f{{\bar b}^2}{4\alp (Y\!-Y_0)} \right\}\ ,
\end{equation}
where $$\bar b = b+ 2\sqrt{\frac\alp\mu}\log(\mu (Y\!-Y_0)).$$ One recognize in the 
left-hand of \eqref{modif} the solution of  Eq.\eqref{langevin} reduced to the linear terms. Then  \eqref{modif} expresses the 
universal modification of the large $b$ behavior of the 
amplitude due to the the universal leading-edge structure

A similar study can be made for the stochastic case, $e.g.$ from (\ref{resultsstoch}), or even the explicit formulas 
(\ref{frontU},\ref{eq:sol}). A detailed phenomenological study of all these aspects is deserved, based on the results of the present theoretical paper.

\end{document}